\documentstyle[aps,preprint,tighten,epsfig]{revtex}
\def\beq{\begin{equation}}
\def\eeq{\end{equation}}
\def\bea{\begin{eqnarray}}
\def\eea{\end{eqnarray}} 
\def\beqa{\begin{equation}\begin{array}{l}}
\def\eeqa{\end{array}\end{equation}}
\def\eqlab#1{\label{eq:#1}}

\def\sla#1{#1 \!\!\! \slash}
\def\slap{p \!\!\! \slash}

\def\ga{\gamma} \def\Ga{{\it\Gamma}}
 \def\De{\Delta}
  \def\eps{\epsilon}

\def\si{\sigma}

\def\pa{\partial}

\def\pa{\partial}

\begin{document}
\draft
\preprint{MKPH-T-01-08}
\title{Hadron structure and the limitations of phenomenological models
in electromagnetic reactions}

\author{J.~H.~Koch,$^1$ V.~Pascalutsa,$^2$ and S.~Scherer$^3$}

\address{$^1$National Institute for Nuclear and High Energy Physics
(NIKHEF)\\
P.O.~Box 41882, 1009 DB Amsterdam, The Netherlands\\
and\\
Institute of Theoretical Physics, University of Amsterdam
\\[1ex]
$^2$Department of Physics, Flinders University, Bedford Park, 
SA 5042, Australia \\[1ex]
$^3$Institut f\"ur Kernphysik, Johannes Gutenberg-Universit\"{a}t, 
D-55099 Mainz, Germany}
\date{August 17, 2001}

\maketitle

\begin{abstract}
The description of electromagnetic reactions at intermediate energies,  
such as pion electroproduction or (virtual) Compton scattering, traditionally  
starts from covariant tree-level Feynman diagrams (Born or pole terms). 
Internal hadron structure is included by means of (on-shell) form factors 
in the vertices while free propagators are used. 
To overcome problems with gauge
invariance, simple prescriptions such as choosing   $F_1^V (q^2) = F_{\pi}
(q^2)$ in pion electroproduction or the  ``minimal substitution''  are used.
We discuss the limitations of such approaches by comparing to the most  
general structure of electromagnetic vertices and propagators for pions and  
nucleons. The recipes to enforce gauge invariance are critically examined and
contrasted with the exact  treatment.  
The interplay between off-shell effects and irreducible ``contact'' terms
is demonstrated for real Compton scattering on a pion.
The need for a consistent microscopic treatment of reaction amplitudes  
is stressed.  Shortcomings of minimal substitution are illustrated through an
example in the framework of chiral perturbation theory. 
\end{abstract}

\section{Introduction}

High-precision experiments at intermediate energies with electromagnetic probes
are presently being carried out with the goal to investigate the internal 
structure of the participating hadrons and to probe details of the 
reaction dynamics. 
A recent example is the measurement of pion
electroproduction on the nucleon to determine  
the axial and electromagnetic form factors of the nucleon and the pion,
respectively \cite{Liesenfeld:1999mv,Volmer:2001ek}.
In this paper we will 
point out some of the limitations and shortcomings in the commonly used 
phenomenological approaches to describe these reactions and show how in 
contrast structure and dynamics are addressed consistently in microscopic 
approaches such as chiral perturbation theory.

Most theoretical descriptions of intermediate-energy reactions, 
such as Compton scattering or pion electroproduction on a nucleon, have
dealt with internal hadron structure and the reaction dynamics
in a more global fashion and not used a microscopic model for these
aspects. Instead, a description based on covariant tree-level Feynman
diagrams (Born or pole terms) has been used,
where the free hadron properties are the building blocks. The limitations
of this approach are immediately obvious from this simple observation: 
these reactions necessarily involve intermediate particles that are not 
free, {\it i.e.} are ``off mass shell." For example, two-step processes
on the nucleon, such   as Compton scattering, necessarily involve off-shell
nucleons; in pion photo- or electroproduction on the nucleon, in addition,
there is an off-shell pion.
 
This fact implies that for the interpretation of these reactions we need
more input than just the free, on-shell properties of the participating
particles. It adds complexity to the theoretical description: the general
structure
of the electromagnetic vertex of a given hadron, for example, is more
complicated and the associated form factors will depend on more scalar
variables than in the free 
case \cite{Bincer:1960tz,Naus:1987kv,Tiemeijer:1990zp}. 
Nevertheless, encouraged by theorems which
express reaction amplitudes in certain low-energy limits in terms of on-shell
properties of the participating particles, it has been customary for decades to
describe two-step processes on {\it e.g.} nucleons also at intermediate
energies in terms of simple
 ``Born amplitudes,'' which only involve on-shell
information in vertices and
 propagators. 

\begin{figure}[ht]
\vspace{0.5cm}
\epsfxsize=7cm
\centerline{\epsffile{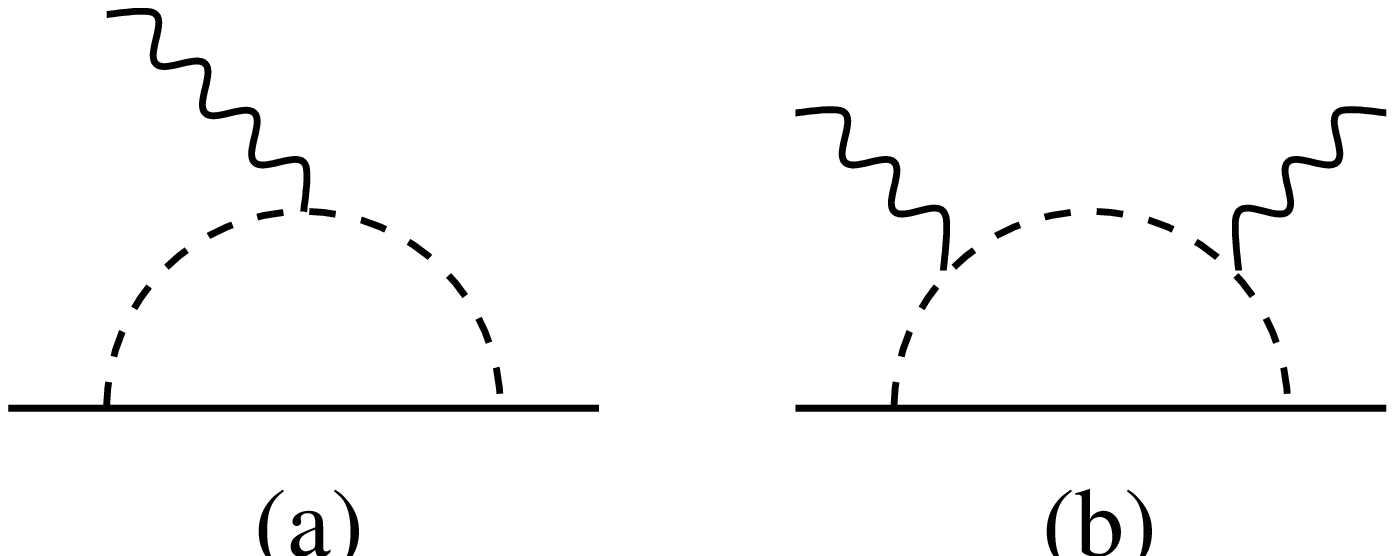}}
\vspace{0.5cm}
\caption[]
{\label{diagramnucleonemvertex:fig}
One-particle-irreducible contributions to the electromagnetic vertex of
a nucleon (a) and to the Compton-scattering amplitude (b).}
\end{figure}

Interestingly enough, the articles deriving these low-energy theorems 
\cite{Kroll:1954xx,Gell-Mann:1954kc}
were the first to point out the more complex structure of vertices and
propagators in the general case at higher energies. What they also emphasized 
was that the internal structure of the hadrons necessarily leads to new 
classes of amplitudes. For example, a diagram that in a given model 
contributes to the electromagnetic structure of a nucleon, will give rise to 
irreducible two-photon amplitudes for Compton scattering 
(see Fig.\ \ref{diagramnucleonemvertex:fig}). 
However, these aspects are absent in the widely used Born-term 
description of Compton scattering or pion photo- and electroproduction.

An objective indication that some important physics is missing in the usual
theoretical approaches is provided by gauge invariance. When form
factors are included in the Born terms, one typically obtains amplitudes 
that are not gauge invariant. Even though the cause of this problem can be
understood and addressed when one consistently deals with the origin of the
internal structure,
 most authors reverted to a variety of {\it ad hoc}
recipes to restore gauge invariance in Born-type amplitudes: terms are added by
hand or form factors are adjusted to yield
 a conserved current. In pion
electroproduction, for example,  a commonly made assumption for this purpose
is that the pion and nucleon 
 isovector form factors are identical 
\cite{Dressler:1988ej,Nozawa:1990gy,Hanstein:1998tp}.
Similary, both the pion-pole term and the nucleon-pole term, which is
needed for gauge invariance, are proportional to the pion form factor 
in the model \cite{Guidal:1997hy,Vanderhaeghen:1998ts} 
used to analyze the data in Ref.\ \cite{Volmer:2001ek}. 
   These assumptions may very well limit the use of this model for extracting 
the pion form factor from experiment.

Another way to deal with this problem are the extended Born terms by
Haberzettl {\em et al.} \cite{Haberzettl:1998eq}. 
They include off-shell modifications at the
hadron vertices in pion photoproduction, consistent with current
conservation. 
 This approach (which has recently been shown to violate crossing
symmetry in its original form \cite{Davidson:2001rk}) 
can, through its additional free parameters, lead to an improved phenomenology.
   However, without a (field-theoretical) microscopic basis it is unclear how
it can be used to reliably extract information about hadron structure. 
   As has been stressed in Refs.\ 
\cite{Scherer:1995aq,Davidson:1996kt,Fearing:1998wq,Fearing:2000fw},
off-shell properties of particles are not observables.
   They are not unique, depending not
only on the microscopic model, but also on the representation one chooses for
the  interpolating fields of the (intermediate) particles
\cite{Scherer:1995aq,Davidson:1996kt,Fearing:1998wq,Fearing:2000fw}.
   In some cases, they can completely be transformed into reaction-specific 
contact terms. 
   When we talk below about ``off-shell effects'' this caveat should be kept 
in mind.

   A direct way to preserve gauge invariance for phenomenological models 
is the minimal substitution. 
It couples a photon to a given charged system,
generating a restricted or 
 ``minimal'' set of terms which guarantee a
gauge-invariant amplitude. 
This prescription has been critically reviewed in
Ref.\ \cite{Heller}.
    It has mainly been applied in pion
photoproduction \cite{Ohta:1989ji}, 
 where the use of a strong pion-nucleon
form 
 factor leads in general to a photoproduction amplitude that is not
gauge 
 invariant. 
The importance of the terms one misses by using this prescription has been 
illustrated in Ref.\ \cite{Bos:1992qs} in the context of
 a simple model for
pion photo- and electroproduction.

In this paper we address these general questions related to the internal
hadron structure in electromagnetic reactions and to gauge invariance. We use 
the pion as the main
example, since the formalism remains simple in comparison to the
nucleon or particles with higher spin, while all important aspects can be
studied. We start in Sec.\ II with a discussion of the features of the general
pion electromagnetic vertex. 
   Many of them have been known for decades 
\cite{Kazes:1959,Barton:1965,Berends:1969bw},
but have largely been ignored in theoretical descriptions of modern 
experiments.
   We examine to what extent this vertex can be obtained through minimal
substitution into the pion self energy for real and virtual photons. 
   We then extend the discussion to the electromagnetic vertex of the nucleon.
   In Sec.\ III
we investigate Compton scattering off a pion as a protoptype for a two-step
process. Special attention is paid to the question how ``off-shell effects''
can contribute to the amplitude. We then contrast the general Compton 
scattering amplitude for a pion with the amplitude generated through minimal 
substitution.

Clearly the most sensible way to address the physics of hadron structure and 
reaction mechanism in a consistent fashion, while avoiding the complications 
and recipes mentioned above, is to work in the framework of a well-defined 
(effective) field theory, such as chiral perturbation theory (ChPT). 
In Sec.\ IV we 
illustrate this by looking at the pion electromagnetic form factor to one loop 
in the framework of ChPT and investigate if the minimal 
substitution recipe can be applied on the level of the Lagrangian itself. 
General conclusions are presented in Sec.\ V.

\section{Electromagnetic
Vertex of Composite Particles: General Features, the Minimal
Substitution and other recipes} 

In this section we discuss the most general structure of the
electromagnetic vertex of spin-$0$ and spin-$\frac{1}{2}$ particles
and contrast it with the vertex of the free particles. We show what
restrictions the Ward-Takahashi (WT) identity 
\cite{Ward:1950xp,Takahashi:1957xn}
imposes and stress
how a consistent treatment of propagator and vertex structure
makes it unnecessary to interrelate pion and nucleon form factors as 
is commonly done.
Finally, we discuss to what extent the minimal substitution 
generates the off-shell vertices needed for the description of the 
electromagnetic interaction with intermediate particles.

\subsection{General features of the pion electromagnetic vertex}

For a spin-$0$ particle (we will for simplicity focus on a
charged pion) Lorentz invariance restricts the form of the 
electromagnetic vertex operator to\footnote{We use the following conventions:
$\hbar=c=1, e>0, e^2/(4\pi)\approx 1/137$, 
$g^{\mu\nu}=\mbox{diag}(1,-1,-1,-1)$,
$\si^{\mu\nu}=(i/2)[\ga^\mu,\ga^\nu]$.} 
\beq
\label{eq:genlorentz}
\Ga^\mu (p\,',p) = e\,\left[(p\,' + p)^\mu F(q^2, p\,'^2, p^2)
+  \ q^\mu \, G(q^2, p\,'^2, p^2)\right]\;,
\eeq
where $p$ and $p\,'$ are the initial and final 
momenta of the meson, respectively, and the photon momentum is
\beq
q= p\,'- p \; . 
\eeq
The neutral pion is its own antiparticle and, because of $C$ invariance, has
no electromagnetic form factor even off shell. The functions $F$ and $G$ are 
form factors which in general depend on three scalar variables. 
For simplicity we do not indicate if a form factor refers to
a positive or negative pion.
   Time-reversal invariance  imposes further
constraints on the functions $F$ and $G$,
\begin{equation}
\label{eq:tri}
F(q^2,p\,'^2,p^2)=F(q^2,p^2,p\,'^2),\quad
G(q^2,p\,'^2,p^2)=- G(q^2,p^2,p\,'^2)\;,
\end{equation}
whereas from Hermiticity follows, for $q^2\leq 0$,
\begin{equation}
F(q^2,m^2,m^2)=F^\ast(q^2,m^2,m^2)\;.
\end{equation}
   Assuming that we are dealing with the irreducible electromagnetic vertex
operator, the requirement of gauge invariance yields the WT identity:
\beq 
\label{eq:wti}
q_\mu\Ga^\mu (p\,',p) = e_\pi\left[\,\De^{-1}(p\,') - \De^{-1}(p)\,\right]\;,
\eeq
where $\De (p)$ denotes the dressed, renormalized propagator of the meson
and $e_\pi=e\,\hat{e}_\pi$ ($\hat{e}_{\pi^\pm}=\pm 1$).
Using the general form in Eq.\ (\ref{eq:genlorentz}), 
the WT identity becomes  
\beq
\label{eq:wti0}
(p\,'^2 - p^2)\; F(q^2,p\,'^2, p^2) + q^2 \;  
G(q^2, p\,'^2, p^2) =  
\hat{e}_\pi\left[ \De^{-1}(p\,') - \De^{-1}(p)\right] \; ,
\eeq
providing a relation between the functions $F$ and $G$ and the 
propagator $\De$ of the meson.
   From Eq.\ (\ref{eq:tri}) or the WT identity it can be easily seen that 
\beq 
\label{eq:G0}
G(q^2, p^2,p^2)=0 \; ,
\eeq
{\it i.e.} this function vanishes when the invariant masses of the initial
and final meson are equal. This implies that $G$ can be written as 
\beq
\label{eq:condG}
G(q^2, p\,'^2, p^2) = (p\,'^2 - p^2)\; g(q^2, p\,'^2, p^2) \; ,
\eeq
where the function $g$ is nonsingular at $p\,'^2 = p^2$.

It is instructive to use the above result to rewrite 
the general electromagnetic vertex of the scalar particle as 
\beq
\label{eq:pifg}
\Ga^\mu (p\,',p) = e \,\left[ (p\,' + p)^\mu  f(p\,'^2, p^2)
+ \left(q^\mu q^\nu - g^{\mu\nu} q^2 \right) (p\,' + p)_\nu \;
 g(q^2, p\,'^2, p^2)\right] \; ,
\label{eq:form}
\eeq
where $f$ is defined as
\bea
\label{eq:deriv}
f(p\,'^2, p^2) &=& F(q^2, p\,'^2, p^2) +\frac{q^2}{p\,'^2 - p^2} 
G(q^2, p\,'^2, p^2) \nonumber  \\
&=& F(q^2, p\,'^2, p^2) + q^2 \, g(q^2, p\,'^2, p^2) \; .
\eea
By contracting Eq.\ (\ref{eq:pifg}) 
with $q_{\mu}$, we see that the function $f$
is entirely determined through the WT identity and depends
only on {\it two} scalar variables, the invariant masses $p\,'^2$ and $p^2$,
\begin{equation}
f(p\,'^2, p^2) =\hat{e}_\pi\frac{\De^{-1}(p\,') - \De^{-1}(p)}{p\,'^2-p^2} \; .
\end{equation}
The second term in Eq.\ (\ref{eq:pifg}) is separately gauge invariant
and no gauge constraints exist for the function $g(q^2, p\,'^2, p^2)$. 
   Only due the presence of this term does the vertex depend on $q^2$.
   In the framework of an effective Lagrangian, such a term can be understood 
in terms of the gauge-invariant combination
\begin{equation}
\label{Fmunu}
F_{\mu\nu}=-i(q_\mu \epsilon_\nu-q_\nu\epsilon_\mu),
\end{equation}
representing the Fourier component of the electromagnetic field
strength tensor ${\cal F}_{\mu\nu}(x)=\partial_\mu A_\nu(x)
-\partial_\nu A_\mu(x)$ associated with a plane-wave 
(virtual) photon $A_\mu(x)=\epsilon_\mu \exp(-iq\cdot x)$.
In a reaction amplitude, the structure multiplying the function $g$ then
results from the contraction of $q_\mu F^{\mu\nu}$ with the sum of the 
pion momenta $(p\,'+p)_\nu$.
   Thus, for the pion, such a separately gauge-invariant term is 
at least of second order in the photon four-momentum.
   Furthermore, powers of $q^2$, $p\,'^2$, and $p^2$ in 
$g(q^2,p\,'^2,p^2)$ can be thought of as originating in powers of (covariant)
derivatives acting on the field-strength tensor, and the final and initial 
charged-particle fields, respectively.

The fact that the function 
$G(q^2, p\,'^2, p^2)$---or $g(q^2, p\,'^2, p^2)$---is 
crucial for the $q^2$ dependence can already be seen by looking at the WT
identity. The right-hand side of Eq.\ (\ref{eq:wti0}) depends only on $p^2$ and
$p\,'^2$, but not on $q^2$. Hence without the term involving $G$, the form
factor $F$ (and with it the entire vertex) could not depend on $q^2$ 
\cite{Barton:1965}! 

However, it should be stressed that the vanishing of  $G(q^2, p\,'^2,
p^2)$ for  $p\,'^2 = p^2$ according to Eq.\ (\ref{eq:G0}) does not rule
out a $q^2$ dependence of the vertex, since in that
limit 
\beq
\lim_{p\,'^2\rightarrow p^2} \Ga^\mu(p\,',p)=
e \,(p+p\,')^\mu F(q^2,p^2,p^2) \; ,
\eeq
and we have from the WT identity, Eq.\ (\ref{eq:wti0}), 
and Eq.\ (\ref{eq:condG}),
\beq
F(q^2,p^2,p^2)=\hat{e}_\pi\frac{\pa \De^{-1}(p)}{\pa p^2} 
 -q^2 g(q^2, p^2,p^2)\;.
\eeq
This includes of course the on-shell case, $p^2 = p\,'^2=
m^2$, which is characterized by a single on-shell form factor
\beq 
F(q^2) \equiv \left. F(q^2, p^2,p^2)\right|_{p^2 = m^2} \; .
\eeq

The significance of the second term in Eq.\ (\ref{eq:genlorentz}) is not 
clearly 
recognized in much of the literature. This term is often dropped
right from the start. The reason given for ignoring it is that a term in the
vertex proportional to $q_{\mu}$ will
not contribute for virtual photons originating from a conserved current.
However, this argument only works in a covariant gauge, {\it e.g.} in
the Feynman or Landau gauge \cite{Naus:1990,Pollock:1996dz}.
In noncovariant
gauges, such as the Coulomb gauge, the gauge term proportional to $q_{\mu}$
must not be neglected since one otherwise would not arrive at the same physical
amplitude as in other gauges. 
By dropping this term in the vertex, 
one gives away the possibility to freely choose
a convenient gauge.

For the interaction of a pion with a real photon, the term
proportional to $q_{\mu}$ can be ignored when calculating an amplitude since
$\eps\cdot q=0$. Since $q^2 = 0$, one only needs   
\bea
\label{eq:real1}
\Ga^{\mu}_{ q^2 = 0}(p\,', p) &=& e \, (p\,' + p)^{\mu} 
f(p\,'^2, p^2) \\
\label{eq:real2}
 &=&  e \, (p\,'+p)^{\mu} F(0, p\,'^2,p^2) \\
\label{eq:real3}
 &=& e_\pi (p\,' + p)^{\mu}\,\frac{\De^{-1}(p\,')
-\De^{-1}(p)}{p\,'^2-p^2} \; .  
\eea
Thus by using Lorentz symmetry and gauge invariance, the
electromagnetic vertex needed for real photons can be shown to be entirely
determined in terms of the finite derivative of the inverse meson propagator. 
This includes the dependence of the vertex on the off-shell variables $p^2$
and $p\,'^2$. The situation is entirely different for virtual photons. The
function $F$ depends on $q^2$ as well and the term involving $G$ must in
general be kept. 

\subsection{The pion form factor in pion electroproduction}

We now address the question of form factors in pion electroproduction on the 
nucleon. 
In the usual description based on pole terms or Born diagrams, 
the pion-pole term is the only term where the
electromagnetic form factor of the pion appears. Since
this term is not
 separately gauge invariant, gauge invariance of the
total amplitude is achieved through cancellation with other terms. To make 
this cancellation with other diagrams, which involve {\it nucleon} form
factors,
 possible, one is forced to make assumptions about these form
factors,
 namely  
\begin{equation} 
\label{eq:fass}
F_1^V(q^2) = F_{\pi}(q^2) \; , 
\end{equation} 
where $F_1^V$ is the electromagnetic isovector form factor of the nucleon.
 In a similar vein, both the reggeized pion-pole term and the
nucleon-pole term were described through the same form factor, $F_\pi$,
to insure gauge invariance in Refs.\ \cite{Guidal:1997hy,Vanderhaeghen:1998ts},
which was the basis for the form factor determination of Ref.\ 
\cite{Volmer:2001ek}.
Clearly such assumptions, which are commonly made 
(see, {\em e.g.}, Refs.\ \cite{Nozawa:1990gy,Hanstein:1998tp})---unless 
they can be put on a firm physical basis---should be avoided when
extracting detailed information such as a pion form factor from
electroproduction experiments.

\begin{figure}[ht] 
\vspace{0.5cm} 
\epsfxsize=4cm 
\centerline{\epsffile{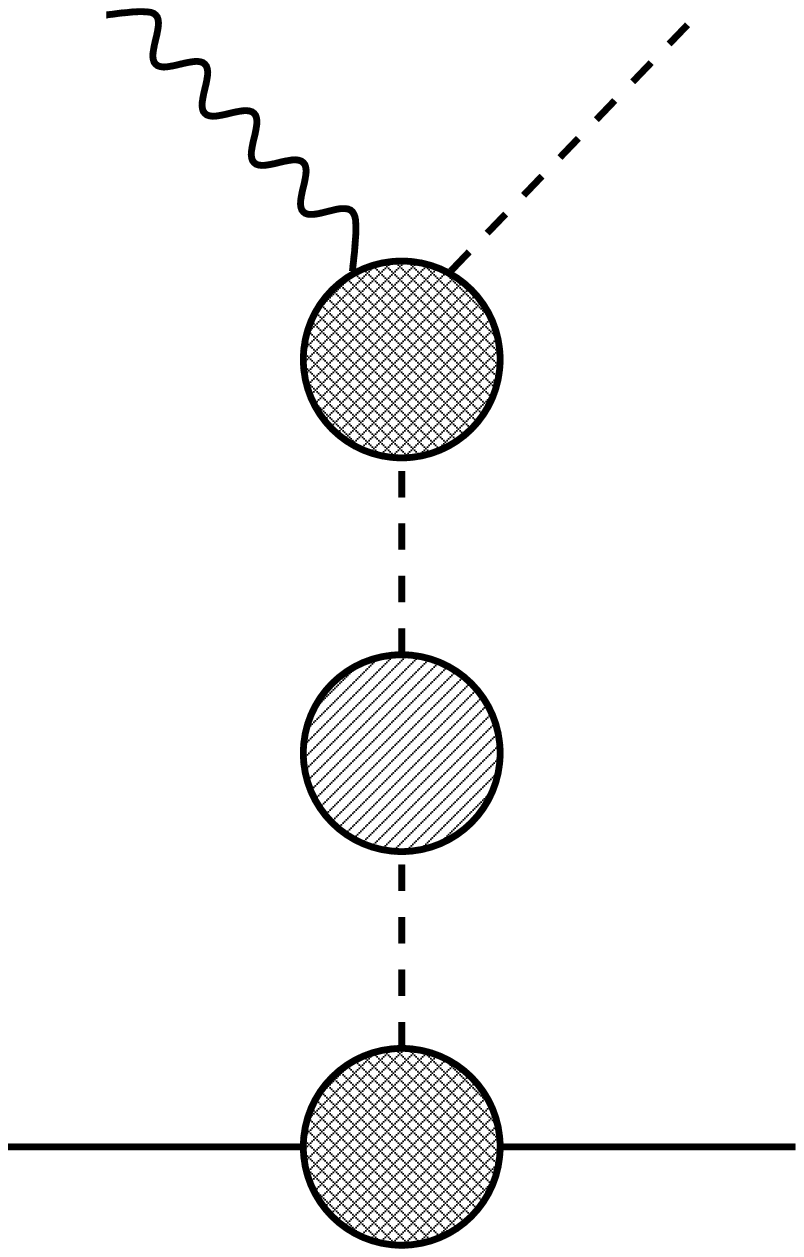}} 
\vspace{0.5cm} 
\caption[] 
{\label{pionpoleterm:fig} 
Pion pole term contributing to charged pion electroproduction.} 
\end{figure}

In contrast, a description that consistently uses a dressed pion propagator 
and vertex in
agreement with the WT identity does not have to make this assumption. The 
intermediate pion in the pion-pole term is half off shell (see Fig.\
\ref{pionpoleterm:fig}). When one  tests gauge invariance and contracts the
pion-pole term with the photon  momentum, it is then easily seen from Eq.\
(\ref{eq:wti}) that, with $p\,'^2 = m^2$ and $\tilde p$ the
momentum of the intermediate
pion, one obtains
\bea
q_{\mu} \, \Gamma^{\mu}(p\,', {\tilde p}) \, \Delta({\tilde p})  &=& e_{\pi} 
 \left[ \Delta^{-1}(p') - \Delta^{-1}({\tilde p}) \right] \, \Delta({\tilde
p}) \nonumber \\
  &=& - \, e_{\pi}  \; .
\eea
The divergence of the pion-pole term, where internal structure is consistently
included in pion vertex and propagator, is thus independent of any detailed
pion properties. Analogous statements
 hold for
the nucleon-pole terms and the mutual cancellation of pion- and nucleon-pole
terms is obtained without any assumptions about the different form factors.
What the WT identity obviously tells us is that we must simultaneously and
consistently deal with both the electromagnetic vertex and the propagator of 
the same hadron to arrive at a gauge-invariant description. 
(Of course, the pion-pole term is
also the only term involving the strong $\pi NN$ form factor for an off-shell
pion. 
We refer to, {\it e.g.}, Refs.\ \cite{Kazes:1959,Naus:1989em} for a discussion 
how to deal consistently with this aspect).

In principle, there is one particular representation, the ``canonical'' form,
where it is possible to work with free electromagnetic form factors and 
propagators. However, the price one pays for this is the presence of a number
of irreducible ``contact'' terms specific for the reaction under study. While
this representation is convenient for general purposes such as the 
derivation of low-energy theorems for (virtual) Compton scattering 
\cite{Lvov:1993fp,Scherer:1996ux},
it has little use in practical applications. Furthermore, the possibility to
move off-shell effects into contact terms makes assumptions about the
analyticity of the amplitude.
   Above production thresholds one will therefore face difficulties in this
respect \cite{Scherer:1995aq}.

\subsection{The minimal-substitution vertex} 

We now compare the general vertex discussed above with the vertex one gets
by minimal substitution, $\Gamma^{\mu}_{\rm MS}$. An extensive review of the 
minimal substitution can be found in Ref.\ \cite{Kondratyuk:1999zm}
and we here only briefly
outline the procedure.
It is obtained by making the
``substitution'' $p \rightarrow p - e_{\pi}\, A$ in $ - \De^{-1}(p)$ and then 
identifying---in general, via a functional derivative---the term linear 
in $A_{\mu}$.  This is
most easily done by expanding the inverse propagator around $p^2 =
m^2$, {\it i.e.}  
\beq
\De^{-1}(p) = \sum\limits_{n=0}^\infty c_n (p^2-m^2)^n, \,\,\,\,\,
c_n = \left. \frac{1}{n!} (\pa/\pa p^2)^n\, \De^{-1}(p)\right|_{p^2= m^2}.
\eeq
The validity of this expansion is implied by the applicability
of the renormalization procedure; in the
on-shell renormalization scheme one has $c_0=0,\, c_1=1$.
In performing the above substitution, we have to keep in
mind that $p_{\mu}$ and the electromagnetic field, $A_{\mu}$, do not commute.
One obtains from the term linear in $A_\mu$ the minimal-substitution vertex
\beq
{\Ga}^\mu_{\rm MS}(p\,', p) = e_\pi  (p\,'+p)^{\mu} \sum_{n=1}^\infty
c_n \sum\limits_{l=1}^n
(p\,'^2 - m^2)^{l-1}\,(p^2- m^2)^{n-l} \;  .  
\eeq
Using the identity 
$$a^n-b^n = (a-b) \sum\limits_{l=1}^n
a^{n-l} \,  b^{l-1} $$ 
we can rewrite the result in terms of the
finite-difference derivative of the inverse propagator:
\bea
{\Ga}^\mu_{\rm MS}(p\,',p) 
&=& e_\pi (p\,' + p)^{\mu}\,  \frac{\De^{-1}(p\,') -
\De^{-1}(p)}{p\,'^2-p^2} \nonumber \\
 &=& e_\pi (p\,' + p)^\mu f(p\,'^2,p^2).
\label{gaminimal}
\eea
   By comparing to the
general form in Eq.\ (\ref{eq:pifg}), one sees that the
minimal-substitution vertex differs in two respects: its operator structure
and the dependence on scalar variables. Minimal substitution generates
the first term $\sim (p\,' + p)^{\mu}$, 
which depends only on $p\,'^2$ and $p^2$.
   It does not produce the second, separately gauge-invariant term, which is 
crucial for having a vertex that depends on the scalar variable $q^2$. 
   In the following we will refer to this second term as ``non-minimal''
in distinction of the minimal term of Eq.\ (\ref{gaminimal}). 
   The minimal-substitution recipe, while producing a vertex 
in general compliance with the WT identity, clearly leads to an
unsatisfactory result for virtual photons.

However, the vertex obtained from minimal substitution into
the  self-energy is precisely the off-shell vertex needed for a {\it real}
photon, ${\Ga}^{\mu}_{{\rm MS}, q^2 = 0}$ in Eq.\ (\ref{eq:real3}),  
\beq
{\Ga}^{\mu}_{\rm MS}(p\,', p) = \Ga^{\mu}_{ q^2 = 0}(p\,', p)\; .
\eeq

Thus for the interaction of a real photon with a pion, the result of minimal
substitution into the self energy  and of a microscopic Lorentz- and 
gauge-invariant calculation are identical, as long as the underlying 
dynamics is
dealt with consistently in the calculation of both the vertex and the self
energy, {\em e.g.} both are calculated to the same order in an expansion
parameter. However, this finding at $q^2 = 0$, where minimal substitution
leads to the same answer as a microscopic calculation, is an exception.  It
is due to the fact that for a scalar particle the WT identity completely
specifies the vertex needed for the interaction with a real photon. For
$q^2\neq 0$ the minimal substitution makes little sense, since it does not
yield a vertex that depends on the scalar variable $q^2$. To obtain such a
dependence we would need a ``non-minimal'' term 
involving the function $g$, which minimal substitution does not generate.

\subsection{Electromagnetic vertex of the nucleon}
In concluding this section, we point out that similar considerations also  
apply to particles with spin 1/2, such as the nucleon  
\cite{Bincer:1960tz,Naus:1987kv,Kazes:1959}. 

The general electromagnetic vertex operator is commonly written in the  
form \cite{Bincer:1960tz} 
\beq 
\Gamma^{\mu} (p\,',p) = e \sum_{i,j = +, -} \Lambda_{i}(p\,') 
\left[ \gamma^{\mu} F_{ij}^1 + i \frac{\sigma^{\mu \nu} q_{\nu}}{2M}\,F_{ij}^2 
+ \frac{q^{\mu}}{M} F_{ij}^3 \right] \Lambda_j(p) \;, 
\eeq  
where the $F_{ij}^{1,2,3}$---the  
twelve form factors that characterize the off-shell  
vertex---are scalar functions of $q^2$, ${p\,'}^2$, and $p^2$, 
and we have used the projection operators 
\beq 
\Lambda_{\pm} (p) = \frac{W \pm \slap}{2W} \; ,\;\;W = \sqrt{p^2} \; . 
\eeq 
For simplicity we do not show a charge (or isospin) index.
To stay close to the discussion of the pion vertex, we choose here 
an alternative form \cite{Barnes:1962}, 
\beq 
\eqlab{Sachs} 
\Gamma^{\mu} (p\,',p) = e \left(1 - \frac{q^2}{4 M^2}\right)^{-1}  
\, \sum_{i,j = +, -} 
\Lambda_{i}(p\,') \left( \frac{P^{\mu}}{M}\,G_{ij}^E + N^{\mu}\, G_{ij}^M  
+ \frac{q^{\mu}}{2M} \, G_{ij}^q  \right) \Lambda_j(p) \;, 
\eeq 
where 
\beq 
N^{\mu} = \frac{\gamma^{\mu} \,\sla{P} \,\sla{q} - \sla{q}\,\sla{P} \, 
\gamma^{\mu}}{4 M^2}   \; , \;\; P = \frac{p\,' + p}{2} \; . 
\eeq 
Note that $N^{\mu}$ satisfies 
\beq 
q \cdot N = 0 \; . 
\eeq 
Just as for the pion, the functions multiplying the 
term proportional to $q^{\mu}$ must vanish for $p^2 = p\,'^2$ and it is 
convenient to introduce [cf.\ Eq.\ (\ref{eq:condG})] 
\beq 
G_{ij}^q (q^2, p\,'^2, p^2) =  (p\,'^2 - p^2) \;  
g_{ij}^q (q^2, p\,'^2, p^2)  \; . 
\eeq 
With this property, it is easily seen that the on-shell matrix element 
of Eq.\ (\ref{eq:Sachs}) yields the well-known parametrization of the free 
current involving the Sachs form factors $G_E$ and $G_M$. 
 
Further restrictions for the vertex arise from the WT identity, 
\beq 
q_{\mu} \, \Gamma^{\mu} (p\,',p) = e_N\, [S^{-1} (p\,') - S^{-1} (p)] \;, 
\eeq 
where $e_N = e \, {\hat e}_N$ is the charge of the nucleon (${\hat e}_p =1, 
\,  {\hat e}_n=0$) and $S^{-1}$ is the inverse 
propagator of the nucleon, which we 
parametrize as   
\beq 
S^{-1}(p)= A(p^2)+ \frac{\slap }{W} \, B(p^2)\; . 
\eeq 
Projecting out terms in the WT identity by using the projection operators 
$\Lambda_{\pm}$, we obtain \cite{Naus:1990} 
\begin{eqnarray} 
\lefteqn{G_{ij}^E (q^2,p\,'^2,p^2) + {q^2} \, g^q_{ij} (q^2,p\,'^2,p^2)=} 
\nonumber\\ 
&&  
\frac{2M}{p\,'^2-p^2}\left(1-\frac{q^2}{4M^2}\right)  
{\hat e}_N 
\left[A(p\,'^2)- A(p^2)+i B(p\,'^2)-j B(p^2) \right]. 
\end{eqnarray} 
We see that for a real photon the four off-shell functions  
$G^E_{\pm \pm} (0,p\,'^2,p^2)$ can 
be expressed through properties of the nucleon propagator.  
Using the above equation to eliminate $G^E_{ij}$, the covariant nucleon vertex
which satisfies the WT  
 identity can be written, in general, in a form
similar to Eq.\ (\ref{eq:pifg}), 
 \bea 
\Gamma^{\mu} (p\,',p) &=& e_N \, (p\,'+p)^\mu\; 
\frac{S^{-1}(p\,')-S^{-1}(p)}{p\,'^2 -p^2}  + e \,  
\left(1-\frac{q^2}{4M^2}\right)^{-1}\nonumber\\ 
&&\times \sum_{i,j = +,-} \Lambda_{i}(p\,') 
\left[ N^{\mu}\, G_{ij}^M(q^2,p\,'^2,p^2) +  (q^\mu\, q^{\nu} - g^{\mu \nu} 
q^2) \,\frac{P_{\nu}}{M} \,g^q_{ij}(q^2,p\,'^2,p^2) \right] \Lambda_{j}(p) \; .
\nonumber \\
&&
\eqlab{WTgaNN} 
\eea 
Just as for the pion, the first part is again independent
of 
 $q^2$ and through the WT identity entirely given in terms of the  
inverse propagator, which fixes its off-shell dependence.  
The $q^2$ dependence resides entirely in the second part, 
which is separately gauge invariant, and contains an off-shell dependence not 
determined by the dressed propagator $S$.  
 
We note in passing that at the tree level with free 
propagators, $S^{-1}=\slap-m$, the first term yields  
\beq 
e_N\frac{(p\,'+p)^\mu \sla{q}}{(p\,'+p)\cdot q} 
\eeq 
rather than the usual $e_N\ga^\mu$. The two expressions can be seen to differ 
by a separately gauge-invariant term. 
 
Analogous to the pion case, minimal substitution will generate only an 
off-shell dependence related to the nucleon propagator.  
As was pointed out in Ref.\ \cite{Kondratyuk:1999zm},  
due to the non-commutativity of the $\gamma$ matrices 
the minimal-substitution result is not unique; different results differ by 
gauge-invariant terms. None of them generates any 
$q^2$ dependence, clearly a shortcoming when applying this recipe to virtual 
photons. 

\section{Real Compton Scattering on a meson}

We now consider real Compton scattering (RCS)
on a meson. To be explicit, we will think of the scattering on a 
$\pi^+$ meson,
\beq
\gamma(q, \epsilon)+\pi^+(p)\to \gamma(q\,', \epsilon\,')+\pi^+(p\,')\;. 
\eeq
The invariant amplitude can be written as
\begin{equation}
{\cal M}=-i \epsilon_\mu\epsilon\,'^\ast_\nu M^{\mu\nu}
=-i \epsilon_\mu\epsilon\,'^\ast_\nu 
(M^{\mu\nu}_{\rm A}+ M^{\mu\nu}_{\rm B})\;,
\end{equation}
where we have divided the total amplitude into the most
general $s$- and $u$-channel meson-pole terms (class A) 
and the rest (class B) \cite{Gell-Mann:1954kc}.\footnote{
The neutral pion is its 
own antiparticle and class A vanishes in this case.}
Class B contains all terms that cannot be reduced to a
form where only a pion propagator connects the two electromagnetic vertices of
the
 pion. It thus contains irreducible diagrams involving the pion as well as
the
 contributions from intermediate states other than the pion.

The ``off-shell effects'' we discussed in the preceeding section are contained
in the class-A terms. In this section, we want to discuss how their presence
affects the general Compton amplitude. Our discussion will mainly be based on
Lorentz
and gauge invariance, but uses also 
the discrete symmetries $P$, $T$, and $C$ as well as crossing symmetries. 
We have to include the 
irreducible meson terms contained in the class B into this discussion. For 
these terms a separate WT identity holds, which relates it to the
electromagnetic meson  vertex. 
As we will see below, the off-shell dependence in one of the two vertices is
cancelled directly by the off-shell meson propagator. This was the starting
point of the work by Kaloshin \cite{Kaloshin:1999ci}, 
who arrived at the conclusion that in the end
{\em all} off-shell effects in class A necessarily cancel.
We will show that this claim is not true.
Finally, we will also compare the general form of the RCS amplitude with
the amplitude obtained from the minimal-substitution
prescription.

\subsection{The general structure of the RCS tensor}

The class-A part of the Compton tensor $M^{\mu\nu}$ has the general form
\cite{Gell-Mann:1954kc,Scherer:1996ux,Fearing:1998gs}
\bea
\label{eq:classA} 
M^{\mu\nu}_{\rm A}&=&\Gamma^\nu(p\,',p\,'+q\,')
\De(p\,'+q\,')\Gamma^\mu(p+q,p) \nonumber \\
&&+ \Gamma^\mu(p\,',p\,'-q) \De(p\,'-q)\Gamma^\nu(p-q\,',p) \; . 
\eea  
The building blocks of this part of the tensor are $\Gamma^\mu$,
the renormalized one-particle-irreducible vertex, and the 
renormalized propagator, $\De$. In other words, this part 
involves quantities that take into account that the intermediate meson is not
on its  mass shell through the dressed propagator $\De$ and the form factor
$F$. We now want to examine to what extent these off-shell effects contribute
to RCS.

For RCS, the external particles are on their mass shell,
$p^2=p\,'^2=m^2$, and for real photons we have
$q^2=q\,'^2=0$.
The class-A contribution therefore reduces to
\bea
\label{eq:claf}
M^{\mu\nu}_{\rm A}&=&e^2\left[(2P^\nu+q^\nu)F(0,m^2,s) \,
\Delta(p\,'+q\,')(2P^\mu+q\,'^\mu)\,F(0,s,m^2)\right. \nonumber \\ 
&&\left.
+ (2P^\mu-q\,'^\mu)F(0,m^2,u) \, \Delta(p\,'-q)(2P^\nu-q^\nu)F(0,u,m^2)
\right]\, , 
\eea
with $2P=p+p\,'$ and the Mandelstam
variables 
\beq
s = (p+q)^2,\quad t=(p-p')^2,\quad u = (p-q\,')^2,\quad
s+t+u=2m^2\;.\nonumber
\eeq
We have dropped all terms proportional to $q^\mu$ and $q\,'^\nu$,
since they do not contribute to the RCS amplitude when contracting
the Compton tensor with the
polarization vectors since $q\cdot\epsilon=q\,'\cdot\epsilon\,'=0$.

The WT identity yields for the form factors in half-off-shell
vertices \cite{Rudy:1994qb}
\begin{equation}
F(0,s,m^2)=\De^{-1}(p\,'+q\,')(s-m^2)^{-1}\; , 
\end{equation}
and
\begin{equation}
F(0,u,m^2)=\De^{-1}(p\,'-q)(u-m^2)^{-1} \; .
\end{equation}
As a result the off-shell dependence of one of the two vertices gets cancelled
by the presence of the dressed propagator  and the
 class-A tensor becomes
\begin{eqnarray}
\label{eq:classarcs1}
M^{\mu\nu}_{\rm A}&=&e^2\left\{ (2P^\nu+q^\nu)
(2P^\mu+q\,'^\mu)\frac{F(0,m^2,m^2)
+[F(0,m^2,s)-F(0,m^2,m^2)]}{s-m^2}\right.\nonumber \\
&&\left.+
(2P^\mu-q\,'^\mu)
(2P^\nu-q^\nu)\frac{F(0,m^2,m^2)+[F(0,m^2,u)-F(0,m^2,m^2)]}{u-m^2}\right\}
\\ 
\label{eq:classarcs2}
&\equiv&M^{\mu\nu}_{\rm pole}+\Delta M^{\mu\nu} \; .
\end{eqnarray}
The total class-A tensor has been split into two parts: The first part involves
on-shell properties only; this part is what one commonly refers to as the 
``pole terms.''
Because of  $F(0,m^2,m^2) = 1$, we have
\begin{equation}  
M^{\mu\nu}_{\rm pole}=  
e^2\left[\frac{(2P^\nu+q^\nu)(2P^\mu+q\,'^\mu)}{s-m^2}
+\frac{(2P^\mu-q\,'^\mu)(2P^\nu-q^\nu)}{u-m^2}\right] \; .  
\end{equation} 
The term $\Delta M^{\mu\nu}$, which is of interest to us, contains all
off-shell contributions of class A. It is proportional to the terms between
square brackets in Eq.\ (\ref{eq:classarcs1}), 
\begin{eqnarray}
\label{deltam}
\Delta M^{\mu\nu}&=&e^2\left[ (2P^\nu+q^\nu)(2P^\mu+q\,'^\mu)h(s)
+(2P^\mu-q\,'^\mu)(2P^\nu-q^\nu)h(u)\right]\nonumber\\
&=&e^2\left\{  4[h(s)+h(u)]P^\mu P^\nu +2 [h(s)-h(u)](P^\mu q^\nu+q\,'^\mu
P^\nu) +[h(s)+h(u)]q\,'^\mu q^\nu\right\}\; ,
\end{eqnarray}
where we have introduced the function
$$
h(z)= \frac{F(0,m^2, z)- F(0,m^2,m^2)}{z-m^2} \; .
$$
   Note that $h(z)$ is analytical for
$z \rightarrow m^2$, because $F(0, m^2, m^2)=1$. 

In order to address the question to what extent the off-shell contribution
$\Delta M^{\mu\nu}$ survives in the total tensor, we first rewrite
$\Delta M^{\mu\nu}$ and $M_B^{\mu\nu}$ in a tensor basis that is
convenient for this purpose. 

The total Compton scattering tensor must be symmetric under
photon crossing,
\beq
q \leftrightarrow - q\,', \; \mu \leftrightarrow \nu\;,
\eeq
which also implies $s\leftrightarrow u$ and $t\to t$.
Since $M_{\rm A}^{\mu\nu}$ is explicitly crossing symmetric, therefore also
$M_{\rm B}^{\mu\nu}$ must be. Similarly, since the pole terms $M^{\mu
\nu}_{pole}$ are crossing symmetric, also $\Delta M^{\mu\nu}$ must have this
property.   
Using furthermore the requirement of invariance under pion crossing in
combination with charge-conjugation invariance,  it is shown in the Appendix
that an appropriate basis for the expansion of
$\Delta M^{\mu\nu}$ and $M_{\rm B}^{\mu\nu}$ is
\bea
M^{\mu\nu}&=&\sum_{i=1}^4{\cal T}_i^{\mu\nu} a_i  \nonumber\\
&=& g^{\mu\nu}\, a_1  +  P^\mu P^\nu\, a_2 \nonumber \\
&+&[4x^2 g^{\mu\nu}-4x(P^\mu q^\nu+q\,'^\mu P^\nu)+ 4 y P^\mu P^\nu]\, a_3
+\frac{1}{2}(yg^{\mu\nu}-q\,'^\mu q^\nu)\, a_4 \; ,\nonumber
\eea
where the coefficients are functions of two kinematical variables, $a_i = a_i
(x^2, y)$,  with
\bea 
x&=&\frac{1}{2}(q+q\,')\cdot P = \frac{1}{4} (s - u)\; , \\
y&=&q\cdot q\,' = - \frac{1}{2}\, t\; .
\eea
   The last two tensors in this expansion are separately gauge invariant, 
while the first two are not. 
Let us first express $\Delta M^{\mu\nu}$ of Eq.\ (\ref{deltam}) in terms of 
this tensor basis, 
\beq
\label{deltam2}
\Delta M^{\mu\nu}=\sum_{i=1}^4 {\cal T}_i^{\mu\nu}\;{\Delta a}_i   \; .
\eeq
By comparing with Eq.\ (\ref{deltam}), we find
\beq
\Delta a_1 = e^2 [F(0,m^2,s)+F(0,m^2,u)-2]\; ,
\eeq
and
\beq
\Delta a_2 = - \frac{2e^2}{x}\,[F(0,u,m^2)-F(0,m^2,s)]\; , 
\eeq
while for the coefficients of the separately gauge-invariant terms we obtain
\begin{eqnarray}
\label{deltaai}
\Delta a_3&=& -\frac{e^2}{2x}\, [h(s) - h(u)]\; ,\nonumber\\
\Delta a_4&=&-2 e^2 [h(s) + h(u)] \; .
\end{eqnarray}

Let us now turn to class B. Using the Ward-Takahashi identity,
Eq.\ (\ref{eq:wti}), 
one can show that gauge invariance of the total tensor, 
\beq
q_{\mu} M^{\mu\nu} = q\,'_{\nu} M^{\mu\nu} = 0,
\eeq
leads to constraints for the class-B part 
\cite{Kazes:1959,Scherer:1996ux,Fearing:1998gs},
\begin{eqnarray}
\label{constraintclassB1}
q_\mu M^{\mu\nu}_{\rm B}&=&e
[\Gamma^\nu(p\,'-q,p)-\Gamma^\nu(p\,',p+q)]\;,\\
\label{constraintclassB2}
q\,'_\nu M^{\mu\nu}_{\rm B}
&=&e[-\Gamma^\mu(p\,'+q\,',p) +\Gamma^\mu(p\,',p-q\,')]\; ,
\end{eqnarray} 
which relates this part of the tensor to the vertex.
   Expanding the class-B tensor with respect to the same basis, 
\beq
\label{deltam2p}
M_{\rm B}^{\mu\nu}=\sum_{i=1}^4 {\cal T}_i^{\mu\nu} b_i,
\eeq
and contracting with $q_\mu$, we obtain
\begin{equation}
\label{qclassB}
q_\mu M^{\mu\nu}_{\rm B}=b_1 q^\nu+x b_2 P^\nu.
\end{equation}
   Using the real photon vertex, Eq.\ (\ref{eq:real2}), we have from
Eq.\ (\ref{constraintclassB1})
\begin{eqnarray}
\label{classBconstraint2}
q_\mu M^{\mu\nu}_{\rm B}
&=&e^2\left\{
2P^\nu[F(0,u,m^2)-F(0,m^2,s)]-q^\nu[F(0,u,m^2)+F(0,m^2,s)]\right\}\;,
\end{eqnarray} 
where we have dropped terms proportional to $q\,'^\nu$
to be consistent with the steps for the class-B parametrization which
led to Eq.\ (\ref{qclassB}).
  By comparing Eqs.\ (\ref{qclassB}) and (\ref{classBconstraint2})
we conclude that
\begin{eqnarray}
\label{A1}
b_1&=&-e^2[F(0,u,m^2)+F(0,m^2,s)]\;,\\ 
\label{A2}
b_2&=& \frac{2e^2}{x}\, [F(0,u,m^2)-F(0,m^2,s)]\;.
\end{eqnarray} 
Since $F(0,u,m^2)- F(0,m^2,s)$ is an odd function of $x$,
there is no singularity for $x \rightarrow 0$ in Eq.\ (\ref{A2}). Using the
second condition of Eq.\ (\ref{constraintclassB2}) 
leads to the same results for $b_1$ and $b_2$ \cite{Scherer:1996ux}.
   Gauge invariance, of course, yields no constraints for $b_3$
and $b_4$.

   When we now combine the contributions from the off-shell tensor, 
$\De M^{\mu\nu}$,
and the class-B part, $M_B^{\mu\nu}$, 
we obtain for the coefficient of the first tensor
 \begin{equation}
\label{DA1+A1}
a_1 = \Delta a_1+b_1=-2e^2\;,
\end{equation}
{\it i.e.}, all off-shell dependence has cancelled and we are left with
a ``seagull term'' which is required by gauge invariance.
Analogously, we find
\begin{equation}
\label{DA2+A2}
a_2 = \Delta a_2+b_2=0 \; ,
\end{equation}
where again the off-shell dependence introduced through the class-A term has
been cancelled.
No such statement can be derived for the coefficients $a_3$ and $a_4$ of the
two separately gauge-invariant tensor structures which are not constrained by
the WT identity.  

In summary, the above procedure has yielded the total result
\begin{eqnarray} 
\label{mmunutotal}
M^{\mu\nu}&=&M^{\mu\nu}_{\rm pole} + \Delta M^{\mu\nu}+M^{\mu\nu}_{\rm B} 
\nonumber
\\
 &=&M^{\mu\nu}_{\rm pole}-2 e^2\, g^{\mu\nu}
+{\cal T}^{\mu\nu}_3\,(\Delta a_3+b_3)
+{\cal T}^{\mu\nu}_4\,(\Delta a_4+b_4)\;.
\end{eqnarray}
Here, 
$$M^{\mu\nu}_{\rm pole}-2 e^2\,g^{\mu\nu}\equiv M^{\mu\nu}_{\rm Born}$$
is the gauge-invariant tensor for a ``point particle,''
commonly denoted as the Born terms.
   It contains the tree-level diagrams of scalar QED without any off-shell
effects in vertices or propagators. 
   The remainder, 
$$(\Delta a_3+b_3){\cal T}^{\mu\nu}_3
+(\Delta a_4+b_4){\cal T}^{\mu\nu}_4
$$
are the separately gauge-invariant parts. It is in these gauge-invariant 
terms where the off-shell effects $\Delta a_3$ and $ \Delta a_4$ enter, 
together with
the class-B contributions.
From the point of view of gauge invariance and the symmetries we have used
above, there is clearly no requirement that all ``off-shell'' contributions 
of the
most general class-A terms are cancelled by class B. It is only true
for the $g^{\mu\nu}$ and $P^\mu P^\nu$ structures, where a cancellation of
off-shell effects {\it must} occur [see Eqs.\ (\ref{DA1+A1}) and
(\ref{DA2+A2})].  This type of cancellation has been observed in other
reactions. Examples are pion photoproduction on the nucleon 
\cite{Workman:1992hr} and NN bremsstrahlung \cite{Scherer:2001hh}.

  As has been discussed by Scherer and Fearing  
\cite{Scherer:1995aq,Fearing:1998wq,Fearing:2000fw},  
there is no absolute and unique meaning to the ``off-shell effects.''  
   By changing the representation of the meson field, 
off-shell effects can be transformed from class-A into class-B terms  
and {\it vice versa}. 
   When carrying out such a transformation to another representation,  
the combined values of $(\Delta a_3+b_3)$ and $(\Delta a_4+b_4)$  
will not change, but the splitting into ``off-shell'' and class-B  
contributions will change.  
   In particular, this means that, {\em e.g.}, the electromagnetic 
polarizabilities one defines from separately 
gauge-invariant terms will, in a general covariant framework,  
receive contributions from both the off-shell behavior of the meson {\em and}  
irreducible class-B terms involving the  
meson or other intermediate particles.  
   A total cancellation of off-shell effects must occur for the two not  
separately gauge-invariant structures ${\cal T}_1$ and ${\cal T}_2$.  
    
   Let us finally comment on Ref.\ \cite{Kaloshin:1999ci}, where it was 
argued that {\em all} off-shell effects will necessarily be cancelled. 
   This total cancellation was inferred from a one-loop calculation  
involving pions and a sigma as dynamical degrees of freedom by investigating 
the s-wave $\pi \sigma$ intermediate state in the direct channel. 
   However, angular momentum conservation forbids a net result for 
$J=0$ in real Compton scattering, no matter where it originates from. 
In other words, the absence of off-shell effects in the s wave can only 
serve as a consistency check of the calculation, but is no proof of a total
cancellation. 
 
\subsection{The minimal substitution RCS amplitude}

We now turn to the possibility of generating a RCS amplitude through minimal
substitution. As we have shown above, the minimal-substitution vertex and the
general vertex are the same for a real photon. Thus the exact class-A term and
the class-A amplitude using the vertex generated by minimal
substitution are the same. However, the class-A term by itself is not gauge
invariant, only the sum of the class-A and class-B amplitudes. One can derive
this class-B amplitude by a microscopic calculation, consistent with the
calculations that yielded the class-A amplitude.
   However, we can also obtain a ``minimal'' class-B term 
$M^{\mu \nu}_{\rm B, MS}$ by performing the minimal substitution in 
$-\Delta(p)$, as in Sec.\ II.C, but now by determinining the 
coefficients of {\em second} order in the electromagnetic field, 
{\em i.e.}, by taking the second functional derivative.
The amplitude corresponding to the tensor   
\beq M^{\mu  \nu}_{\rm MS} = M^{\mu \nu}_{\rm A} + M^{\mu  \nu}_{\rm B, MS}
\eeq 
is then gauge invariant.
The resulting tensor $M^{\mu  \nu}_{\rm B, MS}$ is 
\bea
M^{\mu\nu}_{\rm B, MS} &=& e^2 \left\{-2 g^{\mu\nu} \frac{\De^{-1}(p\,') 
- \De^{-1}(p)}{{p\,'}^2-p^2}\right. \nonumber \\
&&-(2P+q)^\nu (2P+q\,')^\mu \frac{1}{s-p\,'^2}\left[
\frac{\De^{-1}(p+q) - \De^{-1}(p)}{s-p^2} - 
\frac{\De^{-1}(p\,') - \De^{-1}(p)}{p\,'^2-p^2}\right] \nonumber\\
&&\left. - (2P-q\,')^{\mu}(2P-q)^\nu \frac{1}{u-p^2}\left[
\frac{\De^{-1}(p-q\,') - \De^{-1}(p\,')}{u-p\,'^2} - 
\frac{\De^{-1}(p\,') - \De^{-1}(p)}{p\,'^2-p^2}\right]\right\}\,,
\eea 
which can be shown to be crossing symmetric.
With the initial and final meson on mass shell
\bea
M^{\mu\nu}_{\rm B,  MS} &=& e^2\Big\{ -2  g^{\mu\nu} \nonumber \\
&&- (2P+q)^\nu (2P+q\,')^\mu \frac{1}{s-m^2}\left[
F(0,s, m^2) - F(0,m^2,m^2)\right] \nonumber\\
&&-(2P-q\,')^{\mu}(2P-q)^\nu \frac{1}{u-m^2}\left[
F(0,u,m^2) - F(0, m^2, m^2)\right]\Big\}\;.
\eea 
By comparing to Eq.\ (\ref{eq:classarcs1}), 
we see that the last two terms cancel corresponding class-A terms and 
leave us with a class-A tensor for Compton 
scattering from an on-shell particle, $M_{\rm pole}^{\mu \nu}$, 
while the first term is the
contact term one obtains in scalar QED. 
   As a result, the minimal tensor is free of any ``off-shell'' 
properties and identical to the Born tensor,
\beq
M_{\rm MS}^{\mu \nu}  = M_{\rm Born}^{\mu \nu}\; .
\eeq
This result for the minimal-substitution amplitude was already mentioned in
\cite{Nagorny:1999wx} when deriving the VCS amplitude for a
``Dirac proton'' by minimal substitution. 
Again, as for the electromagnetic vertex for a
real photon, this simple result for RCS from a spin-$0$ meson is an exception.

Minimal substitution in other cases {\it does} yield results
with explicit reference to off-shell properties of the particles involved. An
example is the minimal substitution into the $\pi N N$-vertex in pion
photoproduction, as was discussed by Ohta \cite{Ohta:1989ji}. It was
also examined for pion photo- and electroproduction by
Bos {\em et al.} \cite{Bos:1992qs}, who contrasted the 
minimal-substitution amplitude
with the exact result of a model calculation.

\section{Electromagnetic structure in effective field theories}

Clearly a way to deal with particle structure and reaction mechanism in
a consistent way and free of the shortcomings discussed above, is to
start from a microscopic (effective) field theory. This will generate
gauge-invariant amplitudes and take into account vertex structure,
particle self energies {\it etc}. One may be tempted to assume that
use of the minimal substitution on the Lagrangian level is a reliable
procedure. Interestingly enough, this is not as simple and non-minimal terms
involving field-strength tensors play an important role.
 To demonstrate this, we will discuss two different
ways of obtaining 
 the interaction of
a charged pion with an external electromagnetic probe through minimal 
substitution in the context of an effective field theory---namely,
chiral perturbation theory \cite{Gasser:1984yg}.
   In the first procedure we start from an effective field theory 
with a global U(1) symmetry and calculate the renormalized propagator 
of the pion to one loop. 
   We then generate a pion electromagnetic vertex via the ``replacement'' 
$p^\mu \to p^\mu -  e_\pi\, A^\mu$ in the (negative of the) inverse of the 
renormalized propagator of a charged pion. 
  This is analogous to the minimal-substitution procedure to obtain a vertex 
  discussed in Sec.\ II.

   In the second version, we make the minimal substitution on the
Lagrangian level. 
   We extend the same globally invariant effective field theory 
to include the electromagnetic interaction 
by replacing ordinary derivatives by appropriate covariant 
derivatives in the Lagrangian. 
With this Lagrangian we then proceed to calculate the electromagnetic 
vertex to one loop.

  For the first method, we start from a {\em globally}
invariant Lagrangian that contains no coupling to an
electromagnetic field. 
The Lagrangian we use is \cite{Leutwyler:1991mz}
\begin{eqnarray}
\label{leffglobal}
{\cal L}_{\rm eff}^{\rm global}&=& {\cal L}^{\rm global}_2 + 
{\cal L}^{\rm global}_4\;, \nonumber \\
{\cal L}^{\rm global}_2 &=&\frac{F_0^2}{4}\mbox{Tr}\left[\partial_{\mu}U 
(\partial^{\mu}U)^{\dagger}
+m_{\pi,2}^2(U^\dagger+U)\right]\;,\nonumber\\
{\cal L}^{\rm global}_4 &=&
\frac{l_1}{4} \left\{\mbox{Tr}[\partial_{\mu}U (\partial^{\mu}U)^{\dagger}] 
\right\}^2
+\frac{l_2}{4}\mbox{Tr}[\partial_{\mu}U (\partial_{\nu}U)^{\dagger}]
\mbox{Tr}[\partial^{\mu}U (\partial^{\nu}U)^{\dagger}]
\nonumber \\
&&+\frac{l_3 m^4_{\pi,2}}{16}\left[\mbox{Tr}(U^\dagger+ U)\right]^2
-\frac{l_7m^4_{\pi,2}}{16}\left[\mbox{Tr}(U^\dagger-U)\right]^2\;.
\end{eqnarray}
  The Goldstone-boson, {\em i.e.}, pion fields are contained in an 
SU(2)-valued matrix $U$. 
   The constant $F_0$ is the pion-decay constant in the chiral limit,
the $l_i$ are low-energy constants not determined by chiral symmetry.   
Equation (\ref{leffglobal}) generates the most general strong interactions
of low-energy pions at ${\cal O}(p^4)$ in the quark-mass
and momentum expansion including chiral symmetry breaking effects due
to the quark masses.   
   Note that in limit of a vanishing quark mass, Eq.\ (\ref{leffglobal}) is 
equivalent to Eq.\ (2) of Weinberg's original paper on ChPT
\cite{Weinberg:1979kz}.
   
   In the calculations that follow we used the representation 
\begin{eqnarray}
\label{sqrt}
U(x)&=&\frac{\sigma(x)+i\vec{\tau}\cdot\vec{\pi}(x)}{F_0},\quad
\sigma^2(x)+\vec{\pi}^2(x)=F_0^2.
\end{eqnarray}  
The equivalence theorem guarantees that physical observables
do not depend on the specific choice of parametrization of $U$
\cite{Kamefuchi:1961sb}.
   However, separate building blocks such as vertices and propagator, in
general, exhibit different off-shell behavior depending on the choice of
interpolating field \cite{Scherer:1995aq,Fearing:1998wq,Fearing:2000fw}.

\begin{figure}[ht]
\vspace{0.5cm}
\epsfxsize=15cm
\centerline{\epsffile{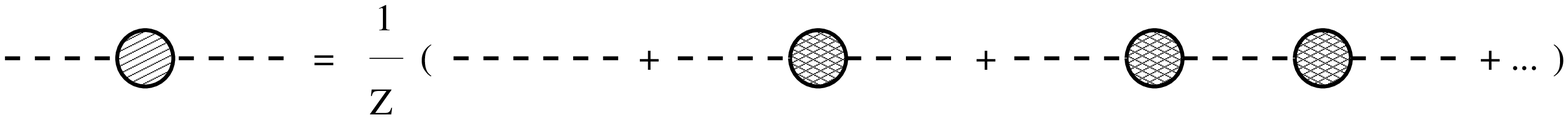}}
\vspace{0.5cm}
\caption[]
{\label{pionpropagator:fig}
Renormalized pion propagator. $Z$ is the wave function renormalization
constant.}
\end{figure}

\begin{figure}[ht]
\vspace{0.5cm}
\epsfxsize=5cm
\centerline{\epsffile{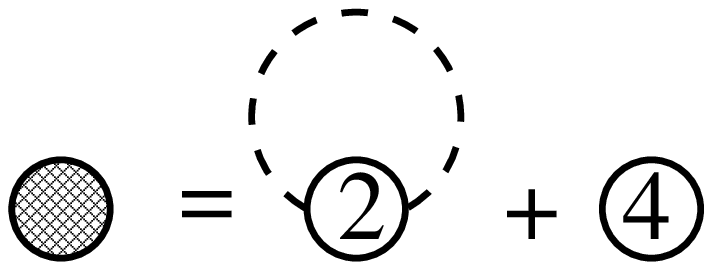}}
\vspace{0.5cm}
\caption[]
{\label{pionselfenergy:fig}One-particle-irreducible pion self energy 
at ${\cal O}(p^4)$. The vertices are derived from
${\cal L}_2$ and ${\cal L}_4$, denoted by 2 and 4 in the interaction
blobs, respectively.}
\end{figure}

   To proceed in analogy with the discussion in Sec.\ II, we first use the
Lagrangian in Eq.\ (\ref{leffglobal}) to determine the renormalized propagator 
at ${\cal O}(p^4)$ in the momentum expansion (see Figs.\ 
\ref{pionpropagator:fig} and \ref{pionselfenergy:fig})
and then generate a pion electromagnetic vertex through minimal substitution.
   To one loop, we obtain for the unrenormalized self energy
\cite{Rudy:1994qb}, 
\begin{equation} 
\label{selfenergy}
\Sigma(p^2) = A + B p^2,
\end{equation}
   where 
\begin{equation}
\label{ab}
A=\frac{3}{2}\frac{m^2_{\pi,2}}{F^2_0}\;  I(m^2_{\pi,2},\mu^2)
+2 \;l_3 \; \frac{m^4_{\pi,2}}{F^2_0} \; ,\quad
B=-\; \frac{I(m^2_{\pi,2},\mu^2)}{F^2_0} \; ,
\end{equation}
and  $I(M^2,\mu^2)$ refers to the dimensionally regularized one-loop integral
\begin{eqnarray}
\label{i}
I(M^2,\mu^2)&\equiv& \mu^{4-d}\int\frac{d^dk}{(2\pi)^d}\frac{i}{k^2-M^2+i0^+}
=\frac{M^2}{16\pi^2}\left[R+\ln\left(\frac{M^2}{\mu^2}\right)\right]
+{\cal O}(4-d) \; ,\\
\label{r}
R&=&\frac{2}{d-4}-[\ln(4\pi)+\Gamma'(1)+1]\; .
\end{eqnarray}
   The renormalized mass and the wave function renormalization constant,
respectively to ${\cal O}(p^4)$ and ${\cal O}(p^2)$, are given by
\begin{eqnarray}
\label{mass}
m^2_{\pi,4} &=& m^2_{\pi,2}(1+B)+A,\\
\label{Z}
Z&=&1+B.
\end{eqnarray} 
At ${\cal O}(p^4)$, the full renormalized propagator is given by
\begin{eqnarray}
\label{propagator}
\Delta_R(p)&=&\frac{1}{Z}\frac{1}{p^2-m^2_{\pi,2}-\Sigma(p^2)+i0^+} 
= \frac{1}{p^2-m^2_\pi +i0^+} \; ,
\end{eqnarray}
where we have replaced the ${\cal O}(p^4)$ expression for the
squared pion mass by the empirical value, the difference being
of ${\cal O}(p^6)$. Minimal substitution into the (negative)
inverse propagator then
leads to the following vertex of a positively charged pion
\begin{equation}
\label{point}
\Gamma^{\mu}(p\,',p) = e (p\,' + p)^{\mu}.
\end{equation}
   Clearly, at ${\cal O}(p^4)$ the minimal-substitution recipe only 
generates the
interaction of a pointlike charged spin-0 field,
without any form factors or $q^2$ dependence.

\begin{figure}[ht]
\vspace{0.5cm}
\epsfxsize=8cm
\centerline{\epsffile{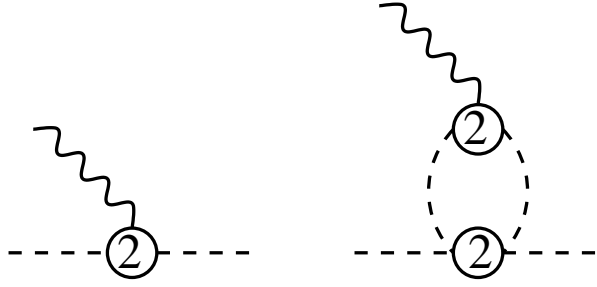}}
\vspace{0.5cm}
\caption[]
{\label{pionemvertexminimal:fig} One-particle-irreducible electromagnetic
vertex at ${\cal O}(p^4)$ obtained from the ``minimal Lagrangian.''
The vertices are derived from ${\cal L}_2$ denoted by 2 in the interaction
blobs.}
\end{figure}

\begin{figure}[ht]
\vspace{0.5cm}
\epsfxsize=4cm
\centerline{\epsffile{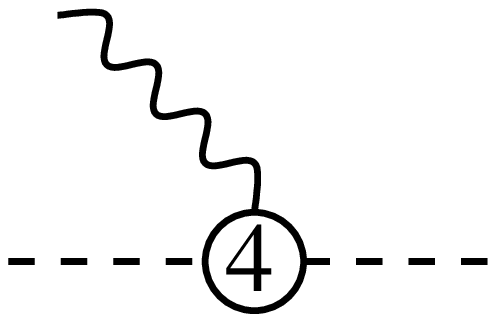}}
\vspace{0.5cm}
\caption[]
{\label{addemvertex:fig} Generic tree-level contribution at ${\cal O}(p^4)$.}
\end{figure}

We now proceed according to the second method to obtain an electromagnetic 
vertex.
We first extend the {\it
Lagrangian} in Eq.\ (\ref{leffglobal}) through minimal substitution,\footnote{
This corresponds to gauging the relevant U(1) subgroup of the global
$\mbox{SU(2)}_L\times\mbox{SU(2)}_R$.}
\begin{equation} 
\label{minsub}
\partial_\mu U\mapsto D_\mu U=\partial_\mu U+\frac{i}{2}eA_\mu[\tau_3,U]\;.
\end{equation}
The calculation of the pion self-energy to order ${\cal O}(p^4)$  based on this
Lagrangian, of course, yields the same result as before, {\it i.e.}
Eqs.\ (\ref{selfenergy}) - (\ref{r}). 
However, a different result  for the electromagnetic vertex to order 
${\cal O}(p^4)$ is obtained using the ``minimal Lagrangian.'' The relevant
diagrams are shown in Fig.\ \ref{pionemvertexminimal:fig}.
Note that the
minimal-substitution procedure of Eq.\ (\ref{minsub}) does {\em not} generate
a tree-level contribution at ${\cal O}(p^4)$, schematically shown in
Fig.\ \ref{addemvertex:fig}, since the candidate terms proportional to
$l_1$ and $l_2$ involve at least four pion fields.
The result for the {\it unrenormalized}
one-particle-irreducible vertex reads 
\begin{eqnarray}
\label{urv}
\Gamma^\mu(p\,',p)&=&e\left\{(p\,'+p)^\mu
\left[1+\frac{I(m^2_{\pi},\mu^2)}{F_\pi^2}\right]
+ (p\,'+p)_\nu \left(q^\mu q^\nu - g^{\mu\nu} q^2 \right)\,
g(q^2,{p'}^2,p^2)\right\}\; ,
\end{eqnarray}
where 
\begin{equation}
\label{f}
g(q^2,{p'}^2,p^2)=\frac{1}{6(4\pi F_\pi)^2}
\left[\ln\left(\frac{m_\pi^2}{\mu^2}\right)+R
+\frac{1}{3}
+\left(1-4\frac{m_\pi^2}{q^2}\right)J^{(0)}\left(\frac{q^2}{m_\pi^2}
\right)\right],
\end{equation}
{\it i.e.} at order ${\cal O}(p^4)$ does not depend on ${p'}^2$ 
and $p^2$. Note that this result still depends on the renormalization scale
$\mu$.
The (infinite) constant $R$ was defined in Eq.\ (\ref{r})
and $J^{(0)}(x)$ is the well-known integral 
\begin{displaymath}
J^{(0)}(x)=\int_0^1 dy\ln[1+x(y^2-y)-i0^+].
\end{displaymath}
All quantities at ${\cal O}(p^4)$ have been replaced by their empirical
values ($F_\pi=92.4$ MeV). 
   
Clearly with this procedure, based on the Lagragian generated through minimal
substitution, we now do obtain a vertex with internal structure. In
particular, through the presence of the separately gauge-invariant term
involving the function $g$ we now do have a dependence on $q^2$. 
The vertex obtained in this way is thus completely
different from the first method, resulting in Eq.\ (\ref{point}).

However, there remains a problem. After multiplication of the
unrenormalized vertex, Eq.\ (\ref{urv}), with the wave function
renormalization constant $Z$, Eq.\ (\ref{Z}), the result still contains
infinite contributions proportional to $R$, even for on-shell pions. Only at
the real-photon point does the result have the correct normalization, 
satisfying the Ward-Takahashi identity in combination with the propagator of
Eq.\ (\ref{propagator}). For $q^2\neq 0$, these infinite terms as well as
terms that depend on the renormalization scale remain.

In order to solve this puzzle, we observe that the most general {\em locally
invariant} effective Lagrangian, which includes the coupling to an
electromagnetic field,  at ${\cal O}(p^4)$ necessarily also
contains ``non-minimal terms'' involving field-strength tensors
such as the $l_5$ and $l_6$ structures of
the effective Lagrangian of Gasser and Leutwyler \cite{Gasser:1984yg},
\begin{eqnarray}
\label{leff}
\label{l4gl}
{\cal L}^{\rm GL}_4 &=&\cdots +
l_5\left[\mbox{Tr}(F^R_{\mu\nu}U F^{\mu\nu}_LU^\dagger)
-\frac{1}{2}\mbox{Tr}(F_{\mu\nu}^L F^{\mu\nu}_L
+F_{\mu\nu}^R F^{\mu\nu}_R)\right]\nonumber\\
&&+i\frac{l_6}{2}\,\mbox{Tr}[ F^R_{\mu\nu} D^{\mu} U (D^{\nu} U)^{\dagger}
+ F^L_{\mu\nu} (D^{\mu} U)^{\dagger} D^{\nu} U]+\cdots.
\end{eqnarray}
   Inserting the relevant expression for the field strength tensors,
\begin{equation}
F^{\mu\nu}_L=F^{\mu\nu}_R=-\frac{e}{2}\tau_3 
(\partial^\mu A^\nu-\partial^\nu A^\mu) \; , 
\end{equation}
the $l_6$ term of Eq.\ (\ref{l4gl}) generates an 
additional separately gauge-invariant contact contribution (see
Fig.\ \ref{addemvertex:fig}) 
\bea
\label{dgammamu}
\Delta\Gamma^\mu(p\,',p)&=& e\left[
- (p\,'+p)^\mu \frac{q^2}{F_\pi^2}\; l_6
+(p\,'-p)^\mu\frac{(p\,'^2-p^2)}{F_\pi^2}\; l_6\right] \nonumber\\
&=& e(p\,'+p)_\nu\, l_6\, \frac{q^\mu q^\nu - g^{\mu\nu} q^2}{F_\pi^2}\;.
\eea
Adding the contributions from Eqs.\ (\ref{urv}) and (\ref{dgammamu})
and multiplying the result by the wave function renormalization constant
yields \cite{Unkmeir:2000md}
\begin{equation}
\label{gammamur}
\Gamma^\mu_R(p\,',p)
=e\left\{(p\,'+p)^\mu F(q^2)+(p\,'-p)^\mu \frac{p\,'^2-p^2}{q^2}[
1-F(q^2)]\right\},
\end{equation}
with the form factor $F(q^2)$ at ${\cal O}(p^4)$
(see Eq.\ (15.3) of Ref.\ \cite{Gasser:1984yg})
\begin{equation}
F(q^2)=1-l_6^r\frac{q^2}{F^2_\pi}-\frac{1}{6}\frac{q^2}{(4\pi F_\pi)^2}
\left[\ln\left(\frac{m^2_\pi}{\mu^2}\right)
+\frac{1}{3}
+\left(1-4\frac{m^2_\pi}{q^2}\right)J^{(0)}\left(\frac{q^2}{m^2_\pi}
\right)\right]\;,
\end{equation}
where we introduced $l_6^r=l_6+\frac{R}{96\pi^2}$.
To this order, the
electromagnetic form factor shows no off-shell dependence.   The explicit
dependence on the renormalization scale $\mu$ cancels 
 with a corresponding
scale dependence of the parameter $l_6^r$.
 Clearly, the vertex  in Eq.\
(\ref{gammamur}) and the propagator in Eq.\
 (\ref{propagator}) now satisfy
the Ward-Takahashi identity for arbitrary $q^2$.
          
In the discussion in this section, the first method started out from a
globally invariant effective Lagrangian ${\cal L}^{\rm global}_{\rm eff}$ 
that was used to obtain the renormalized propagator. 
Minimal substitution into the inverse of
this propagator then yields to order ${\cal O}(p^4)$ an electromagnetic vertex
that does not reflect the structure of the pion through a $q^2$ dependence. 
This method only served to illustrate the general discussion in Sec.\ II 
in the context of chiral perturbation theory. 
 
The second method, based on a Lagrangian obtained through
minimal substitution into ${\cal L}^{\rm global}_{\rm eff}$, 
when used to the same order ${\cal O}(p^4)$,
does yield a vertex with $q^2$-dependent form factors. However, we
found that it does not lead
to a  consistent order-by-order renormalizable theory. 
Already at the one-loop level non-minimal terms, {\it i.e.} terms not
generated through minimal substitution in the globally invariant 
Lagrangian, are
mandatory. They lead to separately gauge-invariant contributions to the 
vertex that absorb divergences appearing in one-loop calculations of
electromagnetic processes. Such terms are of course present in the most
general locally invariant Lagrangian.

The above example has shown that a meaningful calculation of the 
electromagnetic
structure of a hadron in an {\it effective} field theory needs 
non-minimal terms. These are not generated by minimal substitution
on the Lagrangian level. The situation would be different in a
truly fundamental theory based on pointlike particles, such
as the standard model, where the underlying electromagnetic coupling 
of bosons and fermions is minimal.

\section{Summary and Conclusions}
Experiments are carried out at the modern electron accelerators
in order to investigate details of the structure of hadrons and
to examine microscopic aspects of the reaction mechanism. A global
understanding of the main features of these reactions, such as pion 
electroproduction or (virtual) Compton scattering on a nucleon, 
is usually achieved in
the context of models that start out from Born-term diagrams, to which 
various improvements are added, such as final-state interactions, resonance
terms and form factors. 
In trying to incorporate the intrinsic structure of the participating hadrons
one typically makes use of a phenomenology taken over from
{\it free} hadrons. 
However, these descriptions necessarily involve intermediate 
particles to which, for example, the photon is coupling. Microscopic models
will yield for their vertex a much more general structure than that
of a free hadron. While similar statements also apply to the strong vertices,
such as the $\pi NN$-vertex,
the insufficient treatment of internal
structure of the electromagnetic vertices leads to an easily noticable 
shortcoming: gauge invariance is not satisfied. It has 
been customary
to deal with this particular problem by invoking {\it ad hoc} recipes. 
Examples are restrictive assumptions about the individual form factors, 
even though these are the objects one wants to study, or the addition of 
extra terms that make the amplitude gauge invariant. 

We started by discussing the general structure of the electromagnetic vertex
of hadrons with spin $0$ and $1/2$, and the restrictions imposed
by covariance and gauge invariance. The most general form of the vertices,
which is manifestly covariant and satisfies the Ward-Takahashi identity,
has a common pattern. 
   First, there is a $q^2$-independent term related 
through gauge invariance to the hadron propagator. Its off-shell dependence
is determined by the dressed propagator. 
   Second, the $q^2$ dependence of the vertex resides entirely in separately
gauge-invariant terms. 
   Their off-shell behavior is not determined by the hadron propagator. 

In phenomenological models a 
frequently used prescription to obtain a gauge-invariant
amplitude is the ``minimal substitution.'' 
Given the self energy of a particle,
it allows one to obtain a vertex that will at least satisfy the
Ward-Takahashi identity.
It is {\it a priori} clear that this method can only yield a very limited
result: it is blind to neutral particles and, for example, cannot
provide an electromagnetic coupling to a neutron. But even where it does 
yield a result, we showed that it is restricted in terms of its
operator structure and the dependence on scalar variables.
An exception is the pion vertex for a real photon, where minimal
substitution into the pion propagator yields the exact result, including 
the dependence on the invariant mass of the off-shell pion.
That is because the non-minimal term in the vertex does not contribute for a
real  photon.  
For virtual photons,  a non-minimal term can contribute, but the
 minimal substitution of course fails to reproduce it. As a result
this method predicts no dependence on $q^2$; this latter statement applies
also to the spin-1/2 electromagnetic vertex.
Nevertheless, minimal substitution has been used
in some instances for virtual photons, see {\it e.g.} \cite{Nagorny:1999wx}.

Another method to enforce gauge invariance in the case of
pion electroproduction is to assume that the pion electromagnetic form
factor and the nucleon isovector form factor are identical,
$F_1^V (q^2) = F_{\pi} (q^2)$, or to assume the pion form factor that
applies to the $t$ channel also can be used for the $s$ or $u$ channel.
This is clearly an assumption one wants to avoid when trying to 
extract the pion electromagnetic form factor from pion electroproduction.
This kind of problem can be expected in all reactions involving different
hadrons.  
We stress that in principle even the
definition of Born terms containing on-shell information, such as
electromagnetic form factors, is not unique. An example is the electromagnetic
vertex of the nucleon parametrized in  terms of the form factors $G_E$ and
$G_M$. It yields of course an on-shell equivalent nucleon current, but
different results when applied off-shell such as in the Born-terms for virtual
Compton scattering. In contrast to the amplitude parametrized with a Dirac-
and Pauli term with the form factors $F_1$ and $F_2$, respectively, the
$G_{E,M}$ version leads to a VCS amplitude that is not gauge invariant unless
specific contact terms are included \cite{Scherer:1996ux}.

After the discussion of the electromagnetic  
vertex of a hadron, we turned to the description of 
two-step processes, using real Compton scattering off a pion as an example.  
 We showed how the general structure of vertices and propagators 
carries through to the final result.  
By starting from the general amplitude, consisting of pole terms and
irreducible contributions, it  was shown how for real photons one arrives
through the requirement of gauge invariance at the Born terms of a point
particle, which are free of any reference to off-shell properties.
The same Born terms were obtained by using minimal
substitution.  

   Off-shell effects were seen to occur in the amplitude through 
separately gauge-invariant terms. 
   Due to photon-crossing symmetry off-shell contributions start appearing in 
the Compton
amplitude only in second order, namely in the polarizabilities. 
(For a discussion of
virtual Compton scattering and systematics of the $q^2$-dependence we refer
to Refs.\ \cite{Scherer:1996ux,Fearing:1998gs}).  It is important that 
these higher-order terms contain  "off-shell behavior" of the pion {\em as well
as}  contributions
 from intermediate states other than a single pion. The
total cancellation of
 off-shell effects observed in the $J=0$ channel  in   Ref.\
\cite{Kaloshin:1999ci} does not prove in general a  complete cancellation of 
off-shell
effects. The observed cancellation only occurs since  due to  angular
momentum conservation there is no net
 $J=0$ contribution. Finally, we
stressed that ``off-shell effects''
 cannot uniquely be associated with
features of a vertex or a propagator; they can, for example,  be shifted from
a vertex to reaction-specific contact terms
of an effective theory.

While the approaches 
based on (extended) Born term amplitudes are often simple and intuitively 
appealing, they 
should be improved in view of  the detailed questions one wants to answer
through the interpretation of
 the measurements with the new generation of
electron accelerators.
 One way to deal consistently with hadron structure in
vertices, propagators,
 irreducible contributions or contact terms,
is by starting out from an (effective) field theory. We demonstrated this
by considering the pion electromagnetic vertex to one loop in chiral
perturbation theory. The Ward-Takahashi identity is satisfied in a 
simple fashion. An important point is that even on this level non-minimal terms
are crucial and the minimal substitution is insufficient. This observation
clearly signals one of the dangers of working with the minimal substitution
on the phenomenological level.

We have here focused mainly on hadron structure in connection with
the electromagnetic interaction. Clearly, analogous considerations 
about the general vertex structure also apply to the strong
vertices. This has received less attention in the literature and
most of the prescriptions invented  for reactions
involving
extended, off-shell hadrons have been designed to insure a 
gauge-invariant amplitude. 
The microscopic field theoretical approaches now being used for strong
interaction processes such as $\pi N$ 
\cite{Gasser:1988rb,Mojzis:1998tu,Fettes:1998ud,Becher:2001hv}
or $NN$ 
\cite{Weinberg:1991um,Ordonez:1996rz,Kaiser:1997mw,Epelbaum:2000dj}
scattering 
provide a firm basis for a meaningful interpretation, since they are
consistently dealing with the internal 
structure, and 
make the introduction of {\it ad hoc} form factors superfluous.

\acknowledgements
   This work was supported by the Deutsche Forschungsgemeinschaft 
(SFB 443), FOM (The Netherlands) and  the Australian Research Council.
\appendix
\section{Appendix}
In general, the Compton tensor of a spin-0 particle can be expressed as 
\begin{eqnarray}
\label{classBpar}
M^{\mu\nu}&=&\sum_{i=1}^{10} M_i \; T^{\mu\nu}_i \nonumber\\
&=& M_1 \, g^{\mu\nu}+M_2 \, P^\mu P^\nu
+M_3 \, (P^\mu q^\nu-q'^\mu P^\nu)
+M_4  \,(P^\mu q^\nu+q'^\mu P^\nu)
+M_5  \,(P^\mu q'^\nu-q^\mu P^\nu)\nonumber\\
&&+M_6 \, (P^\mu q'^\nu+q^\mu P^\nu)
+M_7 \, (q^\mu q^\nu+ q'^\mu q'^\nu)
+M_8  \,(q^\mu q^\nu- q'^\mu q'^\nu)
+M_9  \,q^\mu q'^\nu 
+M_{10} \, q'^\mu q^\nu,\nonumber\\
\end{eqnarray}
where for real Compton scattering, the functions $M_i$ depend on two
kinematical invariants which we choose as
\bea 
x&=&\frac{1}{2}(q+q')\cdot P \; , \\
y&=&q\cdot q'\; .
\eea
Note that for $p^2=p\,'^2=m^2$ one has $x=q\cdot P= q'\cdot P$.
The tensor we are considering is symmetric under
photon crossing, 
\beq
q \leftrightarrow - q', \; \mu \leftrightarrow \nu.
\eeq
This implies for the functions $M_i$ that
\begin{equation} 
\label{photoncrossing}
M_i(x,y)=\pm M_i(-x,y),\quad +:\, i=1,2,3,5,7,9,10,\quad
-:\,i=4,6,8.
\end{equation}
Invariance under pion crossing in combination with
charge-conjugation invariance \cite{Drechsel:1997ag}  
\beq
M^{\mu\nu}(P,q,q')=M^{\mu\nu}(-P,q,q')
\eeq
leads to
\begin{equation}
\label{cpioncrossing}
M_i(x,y)=\pm M_i(-x,y),\quad +:\, i=1,2,7,8,9,10,\quad
-:\,i=3,4,5,6.
\end{equation}
The combination of Eqs.\ (\ref{photoncrossing}) and (\ref{cpioncrossing})
then yields
\beq
M_3=M_5=M_8=0 \; .
\eeq
Furthermore we can extract appropriate powers of $x$ such that the
invariant amplitudes are functions of $x^2$ only.
For real photons, terms proportional to either $q^\mu$ or
$q'^\nu$ can be omitted and finally
the tensor can thus be written as
\beq
\label{classBpar2}
M^{\mu\nu}= c_1\; g^{\mu\nu} +c_2 \; P^\mu P^\nu 
+c_3\;  x(P^\mu q^\nu+q'^\mu P^\nu)
+ c_4\;  q'^\mu q^\nu,
\eeq
where $c_i=c_i(x^2,y)$.
We can re-arrange Eq.\ (\ref{classBpar2}) in terms of two structures
which are not gauge invariant and two which are separately gauge invariant,
\begin{eqnarray}
\label{classBpar3}
M^{\mu\nu}&=&\sum_{i=1}^4 a_i\, {\cal T}_i^{\mu\nu}\nonumber\\
&=&a_1 g^{\mu\nu} + a_2 P^\mu P^\nu \nonumber\\
&+&a_3 [4x^2 g^{\mu\nu}-4x(P^\mu q^\nu+q'^\mu P^\nu)+4 y P^\mu P^\nu]
+a_4 \frac{1}{2}(yg^{\mu\nu}-q'^\mu q^\nu) \; .
\end{eqnarray}
This is the tensor which has been used in our discussion of real 
Compton scattering.



\frenchspacing
\begin{thebibliography}{999}

\bibitem{Liesenfeld:1999mv}
A.~Liesenfeld {\it et al.}  [A1 Collaboration],
Phys.\ Lett.\ B {\bf 468}, 19 (1999).

\bibitem{Volmer:2001ek}
J.~Volmer {\it et al.}  [The Jefferson Lab F(pi) Collaboration],
Phys.\ Rev.\ Lett.\  {\bf 86}, 1713 (2001).

\bibitem{Bincer:1960tz}
A.~M.~Bincer,
Phys.\ Rev.\  {\bf 118}, 855 (1960).

\bibitem{Naus:1987kv}
H.~W.~Naus and J.~H.~Koch,
Phys.\ Rev.\ C {\bf 36}, 2459 (1987).

\bibitem{Tiemeijer:1990zp}
P.~C.~Tiemeijer and J.~A.~Tjon,
Phys.\ Rev.\ C {\bf 42}, 599 (1990).

\bibitem{Kroll:1954xx}
N.~M.~Kroll and M.~A.~Ruderman, 
Phys.\ Rev.\ {\bf 93}, 233 (1954).

\bibitem{Gell-Mann:1954kc}
M.~Gell-Mann and M.~L.~Goldberger,
Phys.\ Rev.\  {\bf 96}, 1433 (1954).

\bibitem{Dressler:1988ej}
E.~T.~Dressler,
Can.\ J.\ Phys.\  {\bf 66}, 279 (1988).

\bibitem{Nozawa:1990gy}
S.~Nozawa and T.~S.~Lee,
Nucl.\ Phys.\  {\bf A513}, 511 (1990).

\bibitem{Hanstein:1998tp}
O.~Hanstein, D.~Drechsel, and L.~Tiator,
Nucl.\ Phys.\  {\bf A632}, 561 (1998).

\bibitem{Guidal:1997hy}
M.~Guidal, J.~M.~Laget, and M.~Vanderhaeghen,
Nucl.\ Phys.\  {\bf A627}, 645 (1997).

\bibitem{Vanderhaeghen:1998ts}
M.~Vanderhaeghen, M.~Guidal, and J.~M.~Laget,
Phys.\ Rev.\ C {\bf 57}, 1454 (1998).

\bibitem{Haberzettl:1998eq}
H.~Haberzettl, C.~Bennhold, T.~Mart, and T.~Feuster,
Phys.\ Rev.\ C {\bf 58}, 40 (1998).

\bibitem{Davidson:2001rk}
R.~M.~Davidson and R.~Workman,
Phys.\ Rev.\ C {\bf 63}, 025210 (2001).

\bibitem{Scherer:1995aq}
S.~Scherer and H.~W.~Fearing,
Phys.\ Rev.\ C {\bf 51}, 359 (1995).

\bibitem{Davidson:1996kt}
R.~M.~Davidson and G.~I.~Poulis,
Phys.\ Rev.\ D {\bf 54}, 2228 (1996).

\bibitem{Fearing:1998wq}
H.~W.~Fearing,
Phys.\ Rev.\ Lett.\  {\bf 81}, 758 (1998).

\bibitem{Fearing:2000fw}
H.~W.~Fearing and S.~Scherer,
Phys.\ Rev.\ C {\bf 62}, 034003 (2000).

\bibitem{Heller}
L.~Heller,  in  {\em The two-body force in nuclei},
edited by S.~Austin and G.~Crawley (Plenum Press, New York, 1972).

\bibitem{Ohta:1989ji}
K.~Ohta,
Phys.\ Rev.\ C {\bf 40}, 1335 (1989).

\bibitem{Bos:1992qs}
J.~W.~Bos, S.~Scherer, and J.~H.~Koch,
Nucl.\ Phys.\ A {\bf 547}, 488 (1992).

\bibitem{Kazes:1959}
E.~Kazes,
Nuovo Cimento {\bf XIII}, 1226 (1959).

\bibitem{Barton:1965}
G.\ Barton, {\em Introduction to Dispersion Techniques in Field
Theory} (Benjamin, New York, 1965) Chap.\ 7.

\bibitem{Berends:1969bw}
F.~A.~Berends and G.~B.~West,
Phys.\ Rev.\  {\bf 188}, 2538 (1969).



\bibitem{Ward:1950xp}
J.~C.~Ward,
Phys.\ Rev.\  {\bf 78}, 182 (1950).

\bibitem{Takahashi:1957xn}
Y.~Takahashi,
Nuovo Cim.\  {\bf 6}, 371 (1957).

\bibitem{Naus:1990}
H.~W.~L.~Naus,
Ph.D.~thesis, Universiteit van Amsterdam, Amsterdam, 1990.

\bibitem{Pollock:1996dz}
S.~Pollock, H.~W.~Naus, and J.~H.~Koch,
Phys.\ Rev.\ C {\bf 53}, 2304 (1996).

\bibitem{Naus:1989em}
H.~W.~Naus and J.~H.~Koch,
Phys.\ Rev.\ C {\bf 39}, 1907 (1989).

\bibitem{Lvov:1993fp}
A.~I.~Lvov,
Int.\ J.\ Mod.\ Phys.\ A {\bf 8}, 5267 (1993).


\bibitem{Scherer:1996ux}
S.~Scherer, A.~Yu.~Korchin, and J.~H.~Koch,
Phys.\ Rev.\ C {\bf 54}, 904 (1996).

\bibitem{Kondratyuk:1999zm}
S.~Kondratyuk and O.~Scholten,
nucl-th/9906044.


\bibitem{Barnes:1962}
K.~J.~Barnes, Phys.\ Lett.\ {\bf 1}, 166 (1962).


\bibitem{Kaloshin:1999ci}
A.~E.~Kaloshin,
Phys.\ Atom.\ Nucl.\  {\bf 62}, 1899 (1999)
[Yad.\ Fiz.\  {\bf 62}, 2049 (1999)].



\bibitem{Nagorny:1999wx}
S.~I.~Nagorny and A.~E.~Dieperink,
Eur.\ Phys.\ J.\ A {\bf 5}, 417 (1999).

\bibitem{Fearing:1998gs}
H.~W.~Fearing and S.~Scherer,
Few Body Syst.\  {\bf 23}, 111 (1998).

\bibitem{Rudy:1994qb}
T.~E.~Rudy, H.~W.~Fearing, and S.~Scherer,
Phys.\ Rev.\ C {\bf 50}, 447 (1994).

\bibitem{Workman:1992hr}
R.~L.~Workman, H.~W.~Naus, and S.~J.~Pollock,
Phys.\ Rev.\ C {\bf 45}, 2511 (1992).

\bibitem{Scherer:2001hh}
S.~Scherer and H.~W.~Fearing,
Nucl.\ Phys.\ {\bf A684}, 499 (2001).

\bibitem{Gasser:1984yg}
J.~Gasser and H.~Leutwyler,
Annals Phys.\  {\bf 158}, 142 (1984).

\bibitem{Leutwyler:1991mz}
H.~Leutwyler,
BUTP-91-26,
{\it Lectures given at 30th Int. Universit\"{a}tswochen f\"{u}r Kernphysik, 
Schladming, Austria, Feb 27 - Mar 8, 1991 and at Advanced Theoretical Study
Inst. in Elementary Particle Physics, Boulder, CO, Jun 2-28, 1991}.

\bibitem{Weinberg:1979kz}
S.~Weinberg,
Physica A {\bf 96}, 327 (1979).

\bibitem{Kamefuchi:1961sb}
S.~Kamefuchi, L.~O'Raifeartaigh, and A.~Salam,
Nucl.\ Phys.\  {\bf 28}, 529 (1961).

\bibitem{Unkmeir:2000md}
C.~Unkmeir, S.~Scherer, A.~I.~L'vov, and D.~Drechsel,
Phys.\ Rev.\ D {\bf 61}, 034002 (2000).


\bibitem{Drechsel:1997ag}
D.~Drechsel, G.~Kn\"{o}chlein, A.~Metz, and S.~Scherer,
Phys.\ Rev.\ C {\bf 55}, 424 (1997).

\bibitem{Gasser:1988rb}
J.~Gasser, M.~E.~Sainio and A.~Svarc,
Nucl.\ Phys.\ {\bf B307}, 779 (1988).

\bibitem{Mojzis:1998tu}
M.~Mojzis,
Eur.\ Phys.\ J.\ C {\bf 2}, 181 (1998).

\bibitem{Fettes:1998ud}
N.~Fettes, U.~Mei{\ss}ner and S.~Steininger,
Nucl.\ Phys.\ {\bf A640}, 199 (1998).

\bibitem{Becher:2001hv}
T.~Becher and H.~Leutwyler,
JHEP {\bf 0106}, 017 (2001).

\bibitem{Weinberg:1991um}
S.~Weinberg,
Nucl.\ Phys.\  {\bf B363}, 3 (1991).

\bibitem{Ordonez:1996rz}
C.~Ordonez, L.~Ray and U.~van Kolck,
Phys.\ Rev.\ C {\bf 53}, 2086 (1996).

\bibitem{Kaiser:1997mw}
N.~Kaiser, R.~Brockmann and W.~Weise,
Nucl.\ Phys.\  {\bf A625}, 758 (1997)

\bibitem{Epelbaum:2000dj}
E.~Epelbaum, W.~Gl\"{o}ckle and U.~Mei{\ss}ner,
Nucl.\ Phys.\  {\bf A671}, 295 (2000).


\end{thebibliography}
\end{document}